\documentstyle[preprint,aps]{revtex}     
\tightenlines
\begin{document}



\title{Ultrathin films of ferroelectric solid solutions under residual depolarizing field}

\author{Igor Kornev, Huaxiang Fu and L. Bellaiche}

\address{Physics Department,
                University of Arkansas, Fayetteville, Arkansas 72701, USA}

\date{\today}

\maketitle

\begin{abstract}
A first-principles-derived approach is developed to study the effects of uncompensated  depolarizing
electric fields on the properties of Pb(Zr,Ti)O$_3$ ultrathin films for different mechanical boundary
conditions. A rich variety of ferroelectric phases and polarization patterns is found, depending on
the interplay between strain and amount of screening of surface charges. Examples include triclinic
phases, monoclinic states with in-plane and/or out-of-plane components of the polarization,
homogeneous and inhomogeneous tetragonal states, as well as, peculiar laminar nanodomains.
\end{abstract}

\pacs{68.55.-a,77.22.Ej,77.80Bh,77.84.Dy}


\narrowtext

\marginparwidth 2.7in
\marginparsep 0.5in

Ferroelectric thin films are of increasing technological interest because of
the need in  miniaturization of devices \cite{Scott}.
An intriguing problem in these films concerns their polarization patterns.
For instance, various patterns have
been recently proposed from measurements, phenomenological theory
and first principles.
Examples are
out-of-plane {\it monodomains} \cite{Tybell,Ghosez,Javier,Streiffer},
180$^{o}$ out-of-plane {\it stripe domains} \cite{Streiffer,Kopal},
180$^{o}$ and 90$^{o}$ multidomains that are oriented {\it parallel} to the
film \cite{Drezner}, and
{\it microscopically-paraelectric} phases \cite{Javier}.
The fact that dramatically different  patterns have been
reported for similar
{\it mechanical} boundary conditions supports a concept discussed in
Refs~\cite{Javier,Mehta,Javier2}, namely
that they arise from different {\it electrical} boundary conditions.
More precisely, real thin films are likely neither in ideal open-circuit conditions ---
for which unscreened polarization-induced surface charges can generate a
large depolarizing
electric field along the growth direction \cite{Meyer,Cohen2}--- nor in ideal short-circuit
conditions --- that are associated with
a vanishing internal field resulting from the full screening of surface charges ---,
but rather
experience a situation in between. As a matter of fact, a reactive atmosphere can lead to a
partial compensation
of surface charges in films with nominal ideal open-circuit conditions \cite{Streiffer}, while
metallic or semiconductor electrodes ``sandwiching''
films do not always provide {\it ideal} short-circuit conditions --- resulting
in a (non-zero) internal field, whose magnitude depends on the element from which
these electrodes are made \cite{Javier}.
The {\it amount} of surface charges' screening in thin films
can thus vary from one experimental set-up to another, possibly generating {\it different}
polarization patterns.

A precise correlation between this amount of screening and
the morphology of the polarization pattern, and how this correlation  depends on {\it mechanical}
boundary conditions,  are still lacking nowadays despite their obvious importance.
Similarly, {\it atomic-scale} details of multidomains - and their formation mechanism -
are mostly unknown in ferroelectric thin films. One may also wonder if
some uncompensated depolarizing fields can
yield unusual polarization pattern, such as monodomains
made of ferroelectric phases that do {\it not} exist in the corresponding
bulk material.
Candidates for these latter anomalous features to occur are films
made of perovskite alloys having a composition lying near their morphotropic
phase boundary (MPB), because of the easiness of rotating their spontaneous polarization
\cite{Nohedareview,JMKiat,PRL5427,IgorPRL2003}.

In this Letter,  we develop a
first-principles-based scheme to investigate the effects of
uncompensated  depolarizing fields on the properties of Pb(Zr,Ti)O$_3$
films near their MPB, for different mechanical boundary conditions.
Answers to the problems summarized above are provided. In particular, we
find a rich variety of ferroelectric phases, including unusual triclinic and monoclinic
states.
We also observe complex nanodomains, and reveal their formation and atomic
characteristics.

Specifically, our Pb(Zr$_{1-x}$Ti$_{x}$)O$_3$ (PZT) thin films
(i) are grown along the [001] direction
(to be chosen along the z-axis); (ii) are ``sandwiched''
between non-polar systems
(mimicking, e.g., air, vacuum, electrodes
and/or non-ferroelectric substrates); (iii) have Pb-O terminated surfaces;
and (iv) have a 50\% overall
Ti composition.
Such low-dimensional structures are modeled by large
periodic supercells that are elongated along the z-direction, and that contain
 a few number of B-layers to be denoted
by $m$ --- with the alloyed atoms being randomly distributed inside each of these
 $m$ planes. (The resulting films thus have a thickness $\simeq$ $4m$ \AA).
Typically,  we use
$10\times 10\times 40$ periodic supercells with $m$ around 5.
The non-polar regions outside the film are thus altogether $40-m$ lattice constant
thick along
the growth direction, which allows well-converged results for the films properties.
The total energy of such supercell is used in Monte-Carlo simulations,
and is written as:
\begin{eqnarray}
   E_{tot} ( \{ { \bf u_{\it i}} \},\{ { \bf v_{\it i}} \}, \eta,
     \{ \sigma_{\it i} \})   =
   E_{Heff} (  \{ { \bf u_{\it i}} \},\{ { \bf v_{\it i}} \}, \eta,\{ \sigma_{\it i} \}) \nonumber \\
    \quad -~~\sum_{i} \beta ~2 \pi~\frac{Z^{2}}{a^3~\epsilon_{\infty}}  <u_{j,z}>_{s} u_{i,z}
   \end{eqnarray}
where $E_{Heff}$ is the (alloy effective Hamiltonian) energy of the
ferroelectric film {\it per se}.
Its expression and first-principles-derived parameters are those given
in Ref. \cite{PRL5427} for bulk PZT.
${\bf u_{\it i}}$ are the local soft modes in unit cells $i$ of the PZT film
--- which are
directly proportional to the electrical polarization and whose components along the
z-axis are denoted as
$u_{i,z}$.
$\{ { \bf v_{\it i}} \}$ are inhomogeneous strain-related variables inside these
films \cite{ZhongDavid}, while
$\{ \eta \}$ is the homogeneous strain tensor. As indicated in Ref. \cite{Karin},
the form of $\{ \eta \}$ (in Voigt notation) is  relevant to two cases of
interest, namely stress-free {\it vs.} epitaxially strained (001) films. In the
former case,  all the components of $\{ \eta \}$ fully relax.
On the other hand, the second
 situation is associated with the freezing of
three components of $\{ \eta \}$ --- i.e., $\eta_6=0$ and $\eta_1=\eta_2=\delta$,
with $\delta$ being the strain resulting
from the lattice mismatch between the film and the substrate --
while the other components relax
during the simulations \cite{Karin}.
$\{ \sigma_{\it i} \}$
characterizes the atomic configuration, i.e.
$\sigma_{\it i}$=+1 or $-1$ corresponds to the
presence of a Zr or Ti atom, respectively, at lattice site $i$ in the PZT film.
The local modes and the inhomogeneous strain-related variables are forced to vanish
in the supercell areas located
outside the PZT films. Therefore, a depolarizing field
implicitly occurs inside the film if this latter has a
component of its polarization along the
growth direction.
The second term of Eq~(1) mimics the effects  of
 an {\it internal} electric field --- that arises from the {\it partial or
full screening } of
polarization-induced charges at the {\it surfaces}, and that is thus
opposite in direction to the {\it unscreened}
depolarizing field --- on the films properties.
This energetic term is dependent on the $Z$ Born effective
charge \cite{ZhongDavid}, the $a$ lattice constant,  the
$\epsilon_{\infty}$ optical
dielectric constant of Pb(Zr$_{0.5}$Ti$_{0.5}$)O$_3$, and  the average of the z-component of the
local modes centered
at the surfaces (denoted by  $<u_{j,z}>_{s}$, and that is self-consistently updated
during the simulations)  \cite{footnote1}.
This second term is also directly proportional to a $\beta$ parameter
that characterizes the {\it strength}
of the E$_{d}$ {\it total} electric field inside the film. Specifically,
$\beta=0$ corresponds to {\it ideal  open-circuit}
boundary conditions with E$_{d}$ having
its maximum magnitude
(when polarizations lie along the z-axis), while an increase in $\beta$ lowers
this magnitude.

The value of $\beta$ resulting in  a vanishing total internal electric field
is dependent on the supercell geometry, and in particular on the number of its
non-polar layers \cite{Meyer}. This {\it ideal short-circuit} $\beta$ is
denoted as $\beta_{SC}$ in the following.
It is found to be 0.69,
{\it for  stress-free films associated with  $m=7$ and a supercell
periodicity of $40$ lattice constant
along the z-axis}, by fitting
the $T=10\,K$ predictions delivered by Eq~(1) to a single result of Ref.~\cite{Ghosez}
(that constructed a microscopic
Effective Hamiltonian
for thin films under stress-free and ideal
short-circuit boundary conditions);
that is,
the polarization in the film layer that is
further away from the surfaces is along the z-axis and has a
magnitude equal to the one in the bulk \cite{footnote2}.
(Note that  Pb(Zr$_{0.5}$Ti$_{0.5}$)O$_3$ {\it bulk} is tetragonal
with an average local mode of 0.1072,
in lattice constant units, at T=10 \,K \cite{PRL5427}).
We found that our resulting $T=10\,K$ layer-by-layer profiles of the
local modes for stress-free ultrathin films, that are under ideal short-circuit electrical
boundary conditions, are remarkably similar to those
of Ref~\cite{Ghosez}. For instance,
the polarization at the surfaces is significantly enhanced with respect to the bulk,
 and increases as the film
thickness decreases. Ref.~\cite{Ghosez} and our proposed scheme
also both predict that the  layers exhibiting the smallest
polarization are those next to the surface layers, and
that the layer that is further away
from the surfaces has a larger polarization for $m=5$ than for
$m=3$ or 7 (thus exhibiting
a non-monotonic behavior versus thickness).

Having demonstrating that our approach can capture subtle details, we now apply it to study the effects
of {\it uncompensated} depolarizing fields on the ground-state of Pb(Zr$_{0.5}$Ti$_{0.5}$)O$_{3}$
 ultrathin films.
The thickness is kept fixed at $m=5$, while different {\it mechanical} boundary conditions
are adopted:
(a) stress-free (Fig~1a); (b) a tensile
strain $\delta= + 2.65 \%$, corresponding to an increase of 0.02 Bohr for the in-plane
lattice constant with respect to Pb(Zr$_{0.5}$Ti$_{0.5}$)O$_3$ bulk (Fig~1b); (c)
a compressive strain  $\delta= -2.65 \%$ (Fig~1c).
More precisely, Figs~1 show the predicted (absolute value of the) Cartesian components $<u_{x}>$, $<u_{y}>$ and
$<u_{z}>$ --- along the [100],  [010] and
[001] directions, respectively --- of the  average of the local mode vectors in the film, as
a function of $\beta$/$\beta_{SC}$ and at T=10\,K.
Figs~1 also display the behavior of $u_{M}$, which is
defined as $u_{M}=\sqrt{<u_x^2+u_y^2+u_z^2>}$ and thus
provides a measure of the {\it local} polarizations.

Under {\it stress-free} conditions, the film has a spontaneous polarization
{\it aligned along the z-axis} for (large) values of $\beta$
that correspond to a screening of at least 98\% of the polarization-induced
surface charges. This results in a tetragonal state to be denoted by T$_z$.
On the other hand, when $\beta$ becomes smaller than $\simeq$ 0.904$\beta_{SC}$, the
internal field along the growth
direction would be too strong to allow an out-of-plane component of the local mode.
As a result, the polarization aligns along an {\it in-plane} $<010>$
direction. The corresponding ferroelectric phase is denoted
as T$_y$. The most striking result for stress-free PZT films is the polarization path
when going from T$_z$ to T$_y$.
As $\beta$/$\beta_{Sc}$ decreases
from 98\% to 90.4\%, the polarization continuously rotates and passes through
{\it three low-symmetry} phases:
a so-called monoclinic M$_A$ state \cite{DavidMorrel} --- occurring for
$0.932 \leq \beta/\beta_{Sc} \leq 0.98$,
and for which $<u_y>$ and $<u_x>$ are nonzero, equal to each other and smaller
than $<u_z>$;
a triclinic $Tr$ phase, for $\beta$/$\beta_{Sc}$ ranging between 92.2 and 93.2\%,
that is characterized by a local mode with non-zero and different Cartesian components; and
a so-called monoclinic M$_c$ ground-state \cite{DavidMorrel}, when $\beta$/$\beta_{Sc}$ ranges
between 90.4 and 92.2\%,
for which  $<u_x>$ vanishes while $<u_y>$ becomes larger than $<u_z>$.
Interestingly, neither the $Tr$ nor the M$_{c}$ phase exists in the temperature-versus-composition
phase diagram of PZT bulk! Decreasing dimensionality thus leads to unusual phases
because of residual depolarizing fields.
Moreover, all the low-symmetry states of
Fig~1a are only stable for compositions lying near the MPB of bulk PZT since we found
that the M$_A$, $Tr$ and M$_C$ phases disappear in favor of T$_z$ for
higher Ti concentration.

Comparing Fig.~1b with Fig.~1a reveals that films under a
{\it tensile} strain react to
depolarizing fields in a dramatically different way than stress-free films.
In particular,
the end-member phases (T$_z$ and T$_y$)
of Fig~1a both disappear. The reason for the vanishing of the
(ideal-short-circuit-derived) T$_z$ phase
is that tensile strains favor in-plane components
of the local modes, because of the
well-known coupling between strain and polarization \cite{ZhongDavid,Karin,Cohen}. Consequently,
T$_z$
transforms into a
M$_A$ phase for large enough $\eta_1=\eta_2=\delta$ strains. These epitaxial
constraints --- and more precisely
the fact that $\eta_2$ can {\it not} be different than  $\eta_1$ --- generate an
in-plane component of the polarization along
the [100] direction in addition to a larger component along the [010] direction, for small $\beta$. The
(ideal-open-circuit-derived)
T$_y$ state thus becomes a M$_C$ phase when going from stress-free to tensile conditions \cite{footnote3}.

Conversely, a large enough {\it compressive} strain annihilates the (in-plane) $<u_x>$ and $<u_y>$
components of the local mode for {\it any} $\beta$ (see Fig.~1c). Two {\it macroscopically}-different
phases result from this annihilation: a {\it paraelectric} $P$ phase for $\beta$ smaller than
0.822$\beta_{SC}$ and a ferroelectric tetragonal T$_z$ phase for larger $\beta$. T$_z$ can be further
separated into two {\it  microscopically}-different phases.  For $\beta$/$\beta_{SC} \geq 0.884$, the
local polarizations all point along the growth direction and have similar magnitude, since  $<u_z>$ is
nearly equal to $u_{M}$. The resulting state is referred to as  $T^{(h)}_{z}$. On the other hand, when
$\beta$ ranges between 82.2\% and 88.4\% of $\beta_{SC}$, $<u_{z}>$ becomes smaller than $u_{M}$. This
characterizes a locally-inhomogeneous polar state to be denoted by $T^{(i)}_{z}$. The transition
between $T^{(h)}_{z}$ and $T^{(i)}_{z}$ is of first-order, as revealed by the sudden jump of both
$<u_z>$ and $u_{M}$. Another striking feature that can be extracted from Fig.~1c when looking at
$<u_z>$ and $u_{M}$ is that the {\it macroscopically-paraelectric} $P$ phase has a relatively large
magnitude for its local polarizations!

In fact, Figs~1 tell us that the only phases for which
the magnitude of the macroscopic polarization significantly differs
from that of the local polarizations are $T^{(i)}_{z}$
and $P$. Under thermal equilibrium,
complex polarization patterns (if any)
should thus only exist for compressively-strained {\it ultrathin} films, while
large ferroelectric monodomains are likely to occur in ultrathin films
that are under stress-free or tensile mechanical boundary
conditions. In order to provide a detailed
{\it microscopic} insight of compressively-strained thin films,
Figs~2 (a), (b) and (c) display a snapshot
of (very large)
 $24\times 24\times 40$ supercell
simulations yielding a $T^{(h)}_{z}$,  $T^{(i)}_{z}$
and $P$ phase, respectively. Fig~2(a) confirms that $T^{(h)}_{z}$ is locally homogenous.
$T^{(h)}_{z}$ is likely the phase observed
in Ref \cite{Tybell} since this latter also occurs for compressive strain
and large screening of the surface charges. More striking, Fig~2(b) reveals that
$T^{(i)}_{z}$ is characterized by the formation of {\it nanodomains} having local dipoles
that are aligned in an opposite direction with respect to the macroscopic  polarization.
These nanodomains propagate throughout the entire thickness of the film but
 are {\it laterally} confined. We believe that
$T^{(i)}_{z}$ is the phase experimentally seen
 in Ref \cite{Streiffer}, and denoted there as F$_{\gamma}$, since this
latter exhibits features
that are consistent with our inhomogeneous $T^{(i)}_{z}$, namely:
(i) compressive strain conditions and {\it partially} compensated depolarizing fields;
(ii) a non-zero spontaneous polarization along the z-axis; and (iii) broad diffraction peaks.
In the $P$ state (see Fig~2c), the nanodomains have laterally ``percolated'',
resulting in the formation of nanoscale 180$^o$ (out-of-plane)
{\it stripe} domains. Such peculiar multidomains are consistent with the so-called
F$_{\alpha}$ and  F$_{\beta}$ phases observed in Ref~\cite{Streiffer}. Another interesting feature
revealed by Fig.~2(c) --- which is, to the best of our knowledge, the first
atomic-scale picture of nanodomains in ferroelectric thin films --- is that each domain is terminated at
{\it one} surface with local dipoles aligned in
opposite {\it in-plane} directions, while the other surface has {\it out-of-plane} dipoles.
This
contrasts with a commonly accepted picture of {\it out-of-plane} 180$^o$ domains in magnetic films
for which  {\it both} surfaces have {\it in-plane} magnetizations \cite{Landau}.
Finally, Fig.~2c also shows that the predicted period $\Lambda$ of the laminar domains is $\simeq$ 8$a_{in}$,
where $a_{in}$ is the in-plane lattice constant. This
remarkably agrees with the measurements of Ref.~\cite{Streiffer} yielding  $\Lambda=37\AA$ for
compressively-strained 20\AA-thick PbTiO$_3$ films.
(Other calculations we performed --- using $n\times n\times 40$ supercells with
$n$=10, 16, 18, 20 and 24 --- confirm that  $\Lambda$ $\simeq$  8$a_{in}$ or 9$a_{in}$)

We thank  D.D. Fong, R. Haumont, J. Junquera, I. Naumov,
S.K. Streiffer and C. Thomson, for having stimulating our interest in
the present topic.
This work is supported by ONR grants N 00014-01-1-0365 and
00014-01-1-0600 and NSF grant DMR-9983678.

\narrowtext

\newpage

\begin{figure}
\caption{Cartesian components of the film average of the local-mode vectors, as a
function of the
$\beta$/$\beta_{Sc}$ parameter in (001) Pb(Zr$_{0.5}$Ti$_{0.5})$O$_3$ ultrathin films having
a $m=5$ thickness, at $T=10K$. Part (a) displays the predictions for stress-film films.
Parts (b) and (c) show
the results for a tensile and compressive
strain of $2.65\%$,  respectively.
The arrows indicate the values of $\beta$/$\beta_{Sc}$ around which phase transitions occur.
The results for $\beta$/$\beta_{Sc}$
down to zero are not shown since they are identical to those associated with
the lowest values of $\beta$/$\beta_{Sc}$ displayed here.
$10\times 10\times 40$ supercells are used. }
\end{figure}

\begin{figure}
\caption{Three-dimensional polarization patterns in
(001) Pb(Zr$_{0.5}$Ti$_{0.5}$)O$_3$ films having
a $m=5$ thickness and that are under a compressive strain of
$-$2.65$\%$, at $T=10K$.
Part (a), (b) and (c) corresponds to a $\beta$/$\beta_{Sc}$  parameter of 94.5\% ($T^{(h)}_{z}$
phase), 87.7\% ($T^{(i)}_{z}$ phase) and 80.8\% ($P$ phase), respectively.
The bottom of Fig.~2c shows the projection of the 3D picture into a X-Z plane.
The red (respectively, blue) color
indicates local dipoles having a positive (respectively, negative)
component along the z-axis. $24 \times 24 \times 40$ supercells are used.}

\end{figure}


\begin{references}
\bibitem{footnote2} $\beta_{Sc}$ for other thickness is derived from the one
associated with
$m=7$ by a rule of three involving the number of non-polar (001) B-layers in
the corresponding supercells.
For instance, $\beta_{Sc}$ is $+0.774$ and $+0.732$ for $m$=3 and 5, when using
$n \times n\times 40$ supercells (with $n$ being an integer).
\bibitem{footnote3} The disappearance of the T$_z$ and the T$_y$ phases
under tensile conditions is a MPB feature, since it does not occur
in PZT ultrathin films having higher Ti composition.
\end{references}
\end{document}